\documentclass[%
 reprint,
 amsmath,amssymb,
 aps,prl,reprint.
]{revtex4-2}

\usepackage{graphicx}
\usepackage{dcolumn}
\usepackage{bm}
\usepackage{hyperref}
\hypersetup{colorlinks=true, citecolor=blue, urlcolor=blue, linkcolor=blue}

\usepackage{mathptmx}
\usepackage{etoolbox}
\usepackage{color}

\newcommand{\blue}[1]{\textcolor{black}{#1}}

\usepackage[export]{adjustbox}

\begin{document}

\preprint{APS/123-QED}

\title{The impact of the interfacial Kapitza resistance on colloidal thermophoresis}

\author{Juan D. Olarte-Plata}
\email{j.olarte@imperial.ac.uk}

\author{Fernando Bresme}%
 \email{f.bresme@imperial.ac.uk}
\affiliation{Department of Chemistry, Imperial College London}
\date{\today}

\begin{abstract}
Thermal gradients impart thermophoretic forces on colloidal particles, pushing colloids towards cold or hot regions, a phenomenon called thermophoresis. Current theoretical relate the Soret coefficient to local changes of the interfacial tension around colloid, which lead to fluid flow around the colloid surface. TheKapitza resistance, a key variable in the description of interfacial heat transport, is an experimentally accessible property that modifies interfacial thermal fields. Here, we introduce a theoretical approach that describes colloid thermophoretic forces by incorporating explicitly Kapitza resistance effects. Our formulation can be used to monitor the dependence of thermophoresis with the interfacial thermal resistance. We show that the Kapitza resistance modifies the thermal field around the colloids, and identify experimental conditions where the Kapitza resistance influences the thermophoretic forces.
We validate our theoretical approach by implementing a non-equilibrium molecular dynamics model of a colloid suspended in a solvent.

\end{abstract}
\maketitle
Thermophoresis describes the motion of colloidal particles in solution. This physical effect was discovered by Ludwig and Soret in the 19$^{th}$ century using alkali halide aqueous solutions~\cite{Ludwig1856,Soret1879}. Thermophoresis is a complex non-equilibrium phenomenon whose explanation has motivated experimental and theoretical workss~\cite{Wiegand_2004,Semenov2004,putnam2005,Wurger2008,Piazza_2008,Duhr19678,galliero2008,brenner2010,Lusebrink_2012,burelbach-zup-2017,arango-restrepo2019,olarte2019,bresme2022}. Following the behavior observed in thermodiffusion, colloids can feature thermophobic/philic behaviour at high/low temperatures. The phobic/philic transition temperature depends on the particle size~\cite{Braibanti-does-thermophoretic-mobility-2008} and screening length, in charged colloids~\cite{Duhr19678}. 
Duhr and Braun introduced the idea of solvation entropy, connecting thermophoresis to interfacial properties~\cite{Duhr19678}. W\"urger~\cite{Wurger2008} developed a hydrodynamic theory where the colloid thermophoresis depends on the solvent and colloid thermal conductivities, the temperature derivative of the solvent-colloid surface energy (surface entropy) and the solvent viscosity. Arango-Restrepo and Rubi~\cite{Arango-soret-faxen-theorem-2019} derived an equation for the Soret coefficient using the Fax\'en theorem. Their equation does incorporate the derivative of the interfacial tension with temperature and the viscosities of the fluid and the particle. 

\blue{Additional work on fluid-solid interfaces has been performed in the context of confined fluids and surfaces \cite{fu-merabia-joly-thermoosmosis2017,Ganti-phys-rev-lett-2017,Cheng-sdighi-majid-jivkov-thermoosmosis-silica-2021,Bjorn-thermoosmosis} focusing on the thermo-osmotic coefficient and following the non-equilibrium thermodynamics~\cite{derjaguin1,derjaguin2,anderson-colloid-transport-1989,kjelstrup-and-bedeaux-book}. The fluid flow induced by thermo-osmosis emerges from stresses along a substrate fluid interface, and it is therefore an interfacial phenomenon. Evidence for fluid flow around nanoparticles has been provided recently using computer simulations~\cite{bresme2022}.}

At the colloid-fluid interface, thermodynamic and transport properties, such as density and thermal conductivity, feature abrupt changes. In the presence of a thermal gradient, these discontinuities give rise to the Kapitza resistance~\cite{Kapitza,Swartz1989} and an interfacial temperature ``jump'' that might influence the thermal field around a colloid, and potentially thermophoresis. The importance of the Kapitza resistance, or its inverse, the Interfacial Thermal Conductance (ITC), $G_K$, will depend on the magnitude of the conductances and, therefore, the colloid-solvent interfacial properties. High ITC will result in small temperature ``jumps'' at the colloid surface, but low ITC might lead to important differences in the interfacial temperatures of the colloid and the solvent. To understand what constitutes high or low ITC, it is instructive to examine experimental studies of hydrophobic and hydrophilic interfaces~\cite{Ge2006}. The reported ITCs vary between $G_K=$50 MW/(K m$^2$), ``low", and 150 MW/(K m$^2$), ``high" ITC. Computer simulations of liquid-vapor interfaces reported even lower ITCs, $\sim$1 MW/(K m$^2$)~\cite{Simon-octane-2004}, or higher $\sim$ 200-300~MW/(K m$^2$)~\cite{Bhattarai-dream-thermal-conductance-2020,olarte-plata-thermal-conductance-water-gold-2022} for gold-water interfaces. At low ITC, $G_K \sim$ 10 MW/(K m$^2$) measurable temperature differences between solvent and colloid, $\Delta T \sim $0.1-1~K, might appear for heat fluxes achievable in micro-fluidics  ($\sim 10^6$~K/m) \cite{Duhr19678} or plasmonic heating (10$^8$ K/m)~\cite{Govorov2006}. 

Here we incorporate ITC effects in the theory of thermophoresis. \blue{Hence, we incorporate in the theory an experimental variable that is becoming increasingly important in the design of thermal management devices, as well as on the interpretation of nanoscale heat transfer experiments.} We show that the ITC modifies the temperature field around a colloid, and the thermophoretic force, particularly at low ITCs. We corroborate the general predictions of the theory using non-equilibrium molecular dynamics simulations.

\begin{figure*}[!ht]
    \centering
    \begin{tabular}{ccc}
   \includegraphics[width=0.33\linewidth]{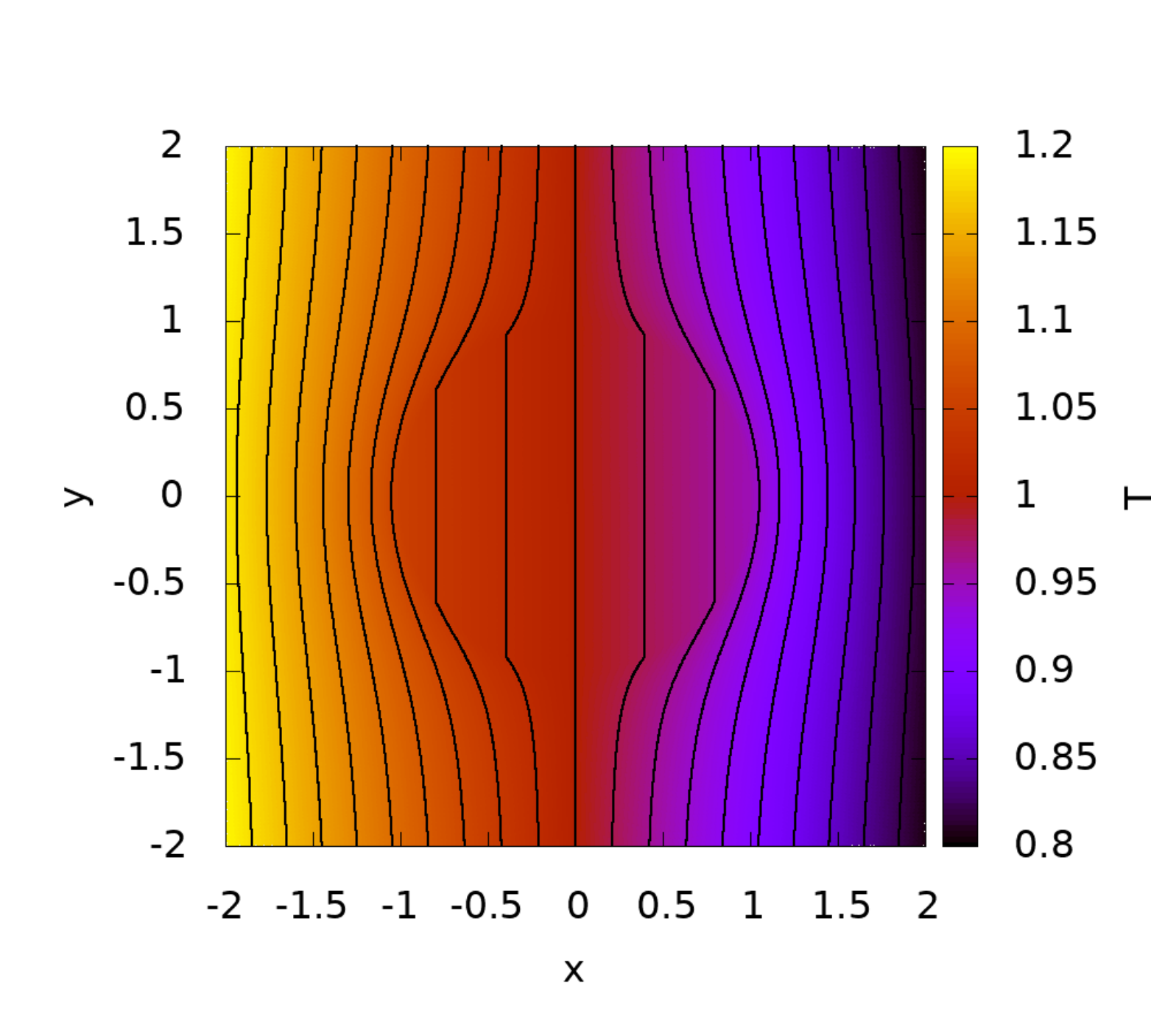} &
    \includegraphics[width=0.33\linewidth]{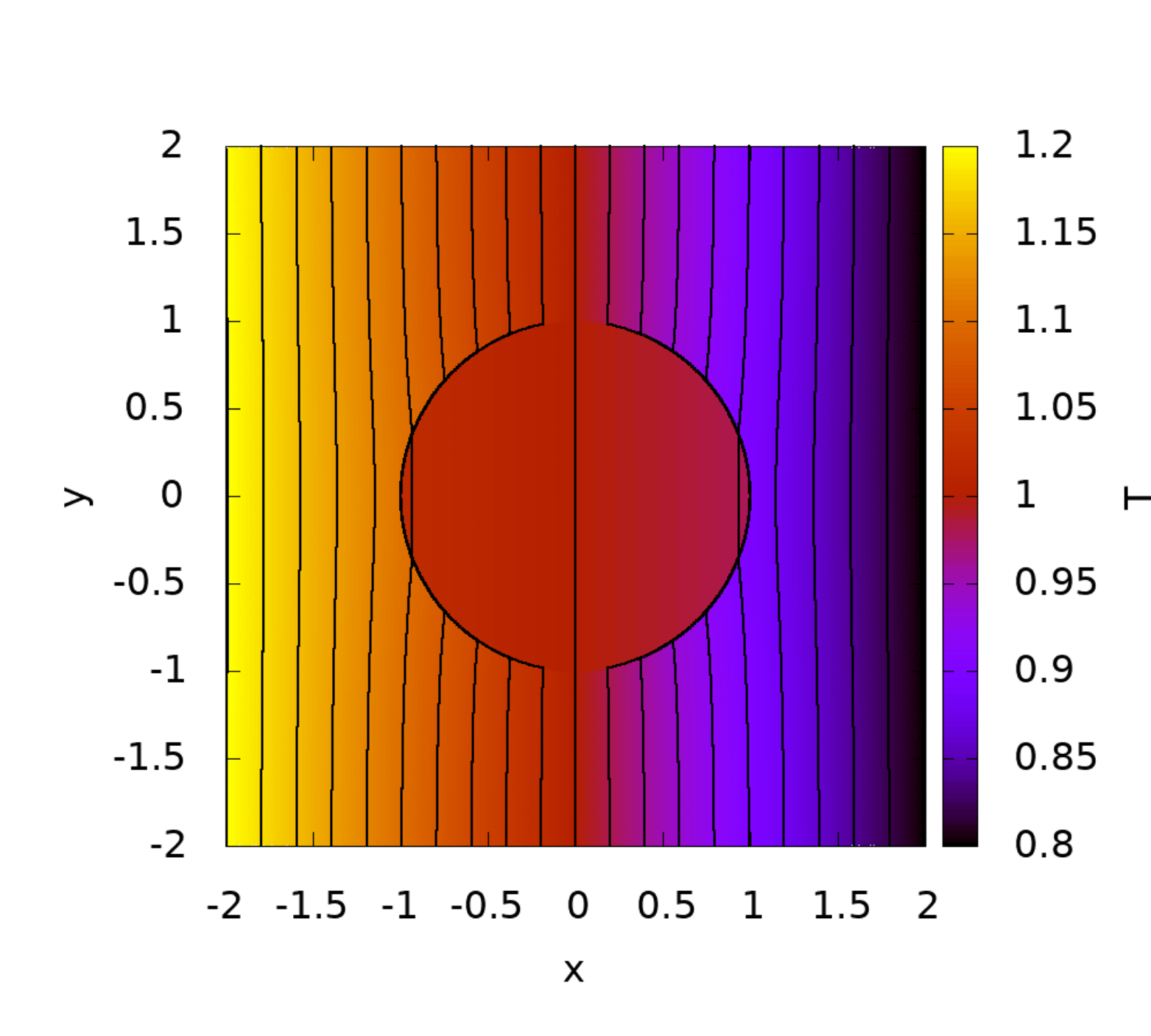} &
     \includegraphics[width=0.33\linewidth]{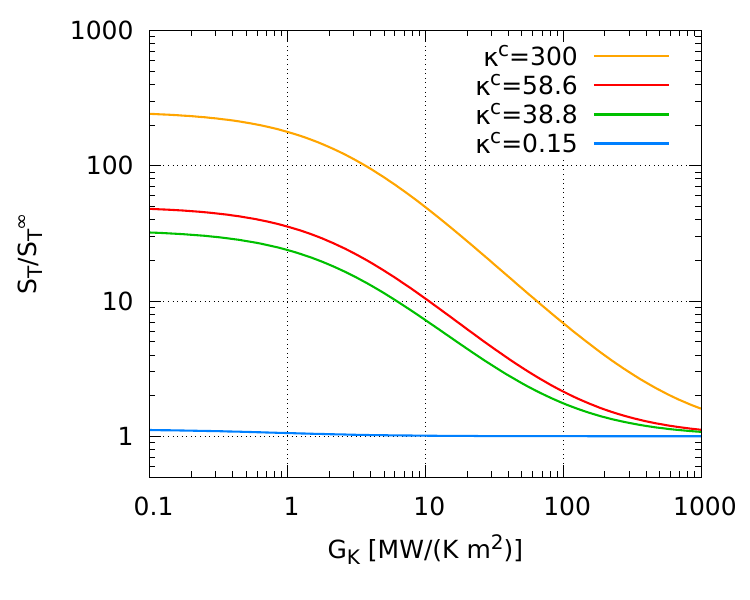} \\
    (a) & (b) & (c)
    \end{tabular}
    \caption{Temperature field around a spherical colloid of radius $R=1.0$, without (a) and with (b) ITC effects. The results in (a) were obtained using Eqs. (4) and (5) in the SI, %(\ref{eq:4}) and (\ref{eq:5}), 
    and those in (b) with Eqs. (\ref{eq:9}) and (\ref{eq:10}). We used $\kappa^s=1.0$, $\kappa^c=4.0$, $G_K=1.0$, average temperature $T_0=1$, and $|\nabla T|=0.1$. These reduced units are compatible with a colloid of radius 250~nm in water,  $\kappa^c / \kappa^s = 4$ ($\kappa^s =0.6$ W/(K m)), $G_k \sim 3$~MW/(K m$^2$). (c) Predicted $S_T$ as a function of $G_K$, for different 
    $\kappa^c$  (in units of W/(K m)) and $\kappa^s=0.6$ W/(K m). The lines represent results obtained with Eq. (\ref{eq:17}). $S_T$ has been divided by the Soret coefficient, $S_T^{\infty}$ at $G_K \rightarrow \infty$, {\it i.e.} when the ITC effects are neglected. The results for $S_T^{\infty}$ were obtained using Eq. (\ref{eq:soret}) and $\alpha$ (Eq. (6) in the SI) instead of $\alpha'$.}
    \label{fig:tmaps}
\end{figure*}

\section{Theoretical model}
Several authors \cite{Wurger2008,Giddings1995,Gaspard2019,Bickel2014} derived equations for the Soret coefficient of spherical colloids by considering the deformation of the temperature field and the ensuing interfacial tension gradient around the colloid, which leads to a Marangoni force. The deformation of the temperature field emerges from the contrast of thermal conductivities of the solvent, $\kappa^{s}$, and that of the colloid, $\kappa^{c}$.
For a spherical colloid immersed in an external temperature field $\nabla T$, of magnitude $|\nabla T|$ in the $\vec{x}$ direction, the temperature profile (in polar coordinates), far away from the particle is:

\begin{equation}
    T^{ext} = T_0 + |\nabla T| \vec{x} = T_0  +  |\nabla T|  r \cos \theta
    \label{eq:1}
\end{equation}

\noindent
where $T_0$ is a reference temperature far from the colloid. Near the colloid, the solvent, $T^{s}(R, \theta)$ and colloid, $T^{c}(R, \theta)$ temperature profiles fulfill the boundary conditions:

\begin{eqnarray}
    T^{s}(R, \theta) & = & T^{c}(R, \theta) \label{eq:boundary1} \\
    \left. \kappa^{s} \dfrac{\partial T^{s}(r, \theta)}{\partial r} \right|_{r=R} & = & \left. \kappa^{c} \dfrac{\partial T^{c}(r, \theta)}{\partial r} \right|_{r=R} \label{eq:boundary2} 
\end{eqnarray}

Following \cite{Giddings1995,Bickel2014,Gaspard2019} the solution of the Laplace equation is:
\begin{eqnarray}
T^{s}(r, \theta) & = & T_{0} + |\nabla T| r \cos \theta \left[1+\alpha\left(\dfrac{R}{r}\right)^3\right] \label{eq:4} \\
T^{c}(r, \theta) & = & T_{0} + |\nabla T| r \cos \theta \left[1+\alpha \right] \label{eq:5}
\end{eqnarray}
\noindent
where $R$ is the colloid radius. The parameter  $\alpha$ quantifies the thermal conductivity contrast between the colloid and the fluid. An explicit equation for $\alpha$ follows from the boundary condition in Eq. (\ref{eq:boundary2}),

\begin{equation}
    \alpha = \dfrac{\kappa^{s}-\kappa^{c}}{2\kappa^{s}+\kappa^{c}}
    \label{eq:6}
\end{equation}

The solvent-colloid interface results in an interfacial thermal conductance that modifies the boundary conditions given by Eqs. (\ref{eq:boundary1}) and (\ref{eq:boundary2}), as the temperature features a discontinuous jump, $\Delta T = \dfrac{J_q}{G_K}$, defined by the heat flux, $J_q$, and the interfacial thermal conductance, $G_K$.
The new boundary conditions, including $G_K$ are,  

\begin{eqnarray}
    T'^{s}(R, \theta) - \dfrac{\kappa^{s}}{G_K} \left. \dfrac{\partial T'^{s}(r, \theta)}{\partial r} \right|_{r=R} & = & T'^{c}(R, \theta) \label{eq:boundary3} \\
    \left. \kappa^{s} \dfrac{\partial T'^{s}(r, \theta)}{\partial r} \right|_{r=R} & = & \left. \kappa^{c} \dfrac{\partial T'^{c}(r, \theta)}{\partial r} \right|_{r=R} \label{eq:boundary4} 
\end{eqnarray}
where the prime indicates the equations include the interfacial thermal conductance effect. The solution of  Eqs. (\ref{eq:boundary3}) and (\ref{eq:boundary4}) gives the corresponding temperature profiles,

\begin{eqnarray}
T'^{s}(r, \theta) & = & T_{0} + |\nabla T| r \cos \theta \left[1+\alpha'\left(\dfrac{R}{r}\right)^3\right] \label{eq:9} \\
T'^{c}(r, \theta) & = & T_{0} + |\nabla T|  r \cos \theta \left[1+\alpha'+\beta' \right] \label{eq:10}
\end{eqnarray}
\noindent 
where $\alpha'$ and $\beta'$ are

\begin{equation}
    \alpha' = \dfrac{\kappa^{s}-\kappa^{c}(1+\beta')}{2\kappa^{s}+\kappa^{c}}
    \label{eq:alphaprimed}
\end{equation}

\begin{equation}
    \beta' = \dfrac{-3\kappa^{s}\kappa^{c}}{G_K R (2\kappa^{s}+\kappa^{c}) + 2\kappa^{s}\kappa^{c}}.
    \label{eq:betaprimed}
\end{equation}
\noindent
For $G_K\rightarrow \infty$, $\beta' \rightarrow 0$ and we recover Eqs.
(\ref{eq:4}) and (\ref{eq:5}). We note that Eqs. (\ref{eq:9}) and (\ref{eq:10}) agree with those derived in reference~\cite{Hasselman1987} to obtain the effective thermal conductivity of composites.

 Fig.~\ref{fig:tmaps} shows the temperature profiles around the colloid with and without interfacial thermal conductance effects. The ITC has a significant impact on the temperature field, and the temperature profile features a discontinuity for the solution that includes the ITC effects. Although the temperature field appears less deformed in the vicinity of the colloid, the ITC leads to different thermal fields far from the colloid surface. The changes in the thermal field influence the thermophoretic force, thermophoretic velocity, and ultimately the Soret coefficient. We now follow the approach introduced in reference~\cite{Wurger2008} to address these changes. We obtain the thermophoretic velocity by considering the surface stress around the colloid. The drift velocity of the particle is given by (see section~1 in the SI for a derivation of the equations given below):

\begin{equation}
\label{eqn:therm-veloc}
    u = -\dfrac{\gamma_{T} R (1+\alpha')}{3 \eta} |\nabla T| 
\end{equation}
\noindent
where $\eta$ is the solvent viscosity and $\gamma_T = d\gamma / dT$ quantifies the change of the interfacial tension, $\gamma$, with temperature. In this theory, the thermophoretic force is a Marangoni force emerging from the surface stress around the colloid due to the temperature gradient. The thermal diffusion coefficient is defined by the 
factor in front of the thermal gradient in Eq. (\ref{eqn:therm-veloc}) 
\begin{equation}
    D_{T} = \dfrac{\gamma_T R (1+\alpha')}{3 \eta}
\end{equation}
\noindent 
which again reduces to the equations reported in previous work~\cite{Wurger2008} when $G_K$ is neglected. 
The Soret coefficient is given by $S_T = D_T/D$, where $D$ is the interdiffusion coefficient approximated for highly diluted suspensions by the colloid diffusion coefficient. For a spherical colloid, $D = k_{B} T /(\xi \pi \eta R)$, where $\xi$ is a numerical parameter accounting for the boundary conditions. The Soret coefficient is given by:

\begin{equation}
    S_T = 4 \pi R^2 \dfrac{(1+\alpha') \gamma_T}{3 k_B T} + \dfrac{1}{T},
    \label{eq:soret}
\end{equation}
\noindent
where the $\alpha'$ parameter considers the different thermal conductivities of the solvent and colloid and the interfacial thermal conductance. The second term of Eq. (\ref{eq:soret}) represents the ideal contribution to the Soret coefficient. The origin of the factor ``4" in Eq. (\ref{eq:soret}) is justified in the derivation included in the SI (see section 2).

The scaling of the Soret coefficient with the square of the colloid radius ($R^2$)  
and the inverse of the temperature is consistent with previous computations of Soret coefficients \cite{Olarte-Plata2020}. However,  $\alpha'$ introduces an additional dependence on the colloid radius through Eq. (\ref{eq:betaprimed}). The solution, including $G_K$ for the solvent and colloid temperature profiles, results in a larger deformation of the temperature field far from the particle (see Figures 1a,b) and a temperature discontinuity at the colloid surface. Based on Eq. (\ref{eq:betaprimed}), we expect that the impact of the interfacial thermal conductance on the Soret coefficient will be stronger when the thermal conductivities of the solvent and the colloid are very different. Figure~1c illustrates the impact of $G_K$ on the predicted Soret coefficient for a colloid of radius $R=250$~nm immersed in water with $\kappa^s=0.6$~W/(K m). For the colloid, we consider a range of thermal conductivities, $\kappa^c=0.15 \cdots 300$~W/(K m). The intermediate thermal conductivities, 58.6 and 38.8~W/(K m), might be representative of the thermal conductivity of a gold nanoparticle coated with a 1 nm or 2 nm alkane passivating layer, with a thermal conductivity for the layer of ~0.45 W/(K m)\cite{olarte-plata-acs-nano2022} (see SI section 7, for more details). We also consider the thermal conductivity of a colloid made of polystyrene with $\kappa^c = 0.15$ W/(K m) as well as high thermal conductivity, of the order of that of bulk gold, $\kappa^c = 300$ W/(K m). The range of ITCs represented in Figure 1c, covers values corresponding to liquid-vapour interfaces ~1 MW/(K m$^2$)~\cite{Muscatello-itc-intrinsic-2017}, and hydrophobic and hydrophilic self assembled monolayers (50-1000 MW/(K m$^2$))~\cite{Ge2006,olarte-plata-acs-nano2022}. For similar solvent and colloid thermal conductivities, the impact of $G_K$ is relatively small, and the results converge to the previous solution that assumes $G_K \rightarrow \infty$. The convergence is much slower when $\kappa^s$ and $\kappa^c$ are very different. Moreover, at low $G_K$, the difference in the Soret coefficient is significant, even for interfacial thermal conductances corresponding to hydrophobic layers, $\sim 50$ MW/(K m$^2$). In summary, our results show that low ITCs can significantly enhance the Soret coefficient when the thermal conductivity of the colloid is much higher than the thermal conductivity of the solvent.

We have shown that the Soret coefficient is proportional to the parameter $1+\alpha'$. We now extend the analysis of the dependence of this parameter on the thermal transport coefficients of the system by introducing the dimensionless quantities $\lambda=\kappa^s/\kappa^c$ and $\epsilon=G_{K} R / \kappa^{s}$. The first quantity is simply the ratio between the thermal conductivities of the solvent and the colloid, while the latter relates the interfacial thermal resistance of the interface with that of the fluid. Using these definitions, we can express the parameter $1+\alpha'$ as:

\begin{equation}
    1+\alpha' = \dfrac{3 \lambda + \dfrac{3}{\epsilon(2\lambda+1)+2}}{2\lambda+1}
\end{equation}

The solution when $\epsilon \rightarrow \infty$ converges to $1+\alpha=3\lambda/(2\lambda+1)=3\kappa^s/(2\kappa^s+\kappa^c)$. The ratio between the solutions with and without considering the ITC can be written as:

\begin{equation}
    \dfrac{S_T}{S_T^{\infty}}= \dfrac{1+\alpha'}{1+\alpha} = 1 + \dfrac{1}{\lambda \epsilon (2\lambda+1)+\lambda}
    \label{eq:17}
\end{equation}
\noindent
where $S_T^{\infty}$ is the Soret coefficient for $G_K \rightarrow \infty$.

\begin{figure*}[!ht]
    \centering
    \begin{tabular}{ccc}
       \includegraphics[width=0.33\linewidth,valign=c]{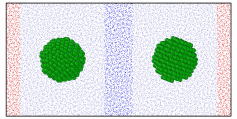} &
        \includegraphics[width=0.33\linewidth,valign=c]{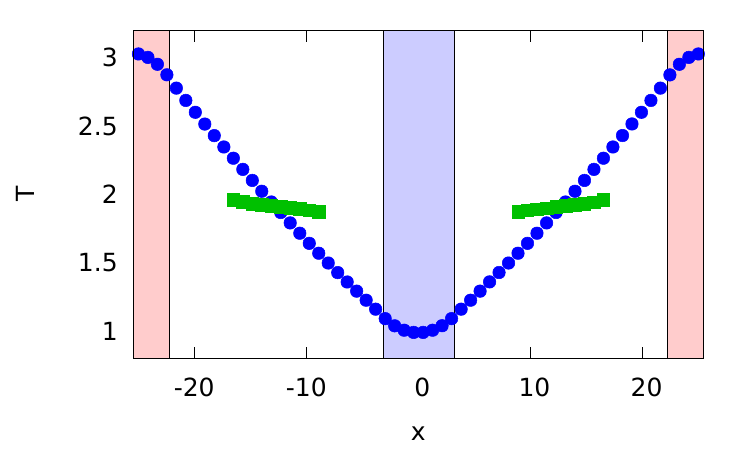} &
      \includegraphics[width=0.33\linewidth,valign=c]{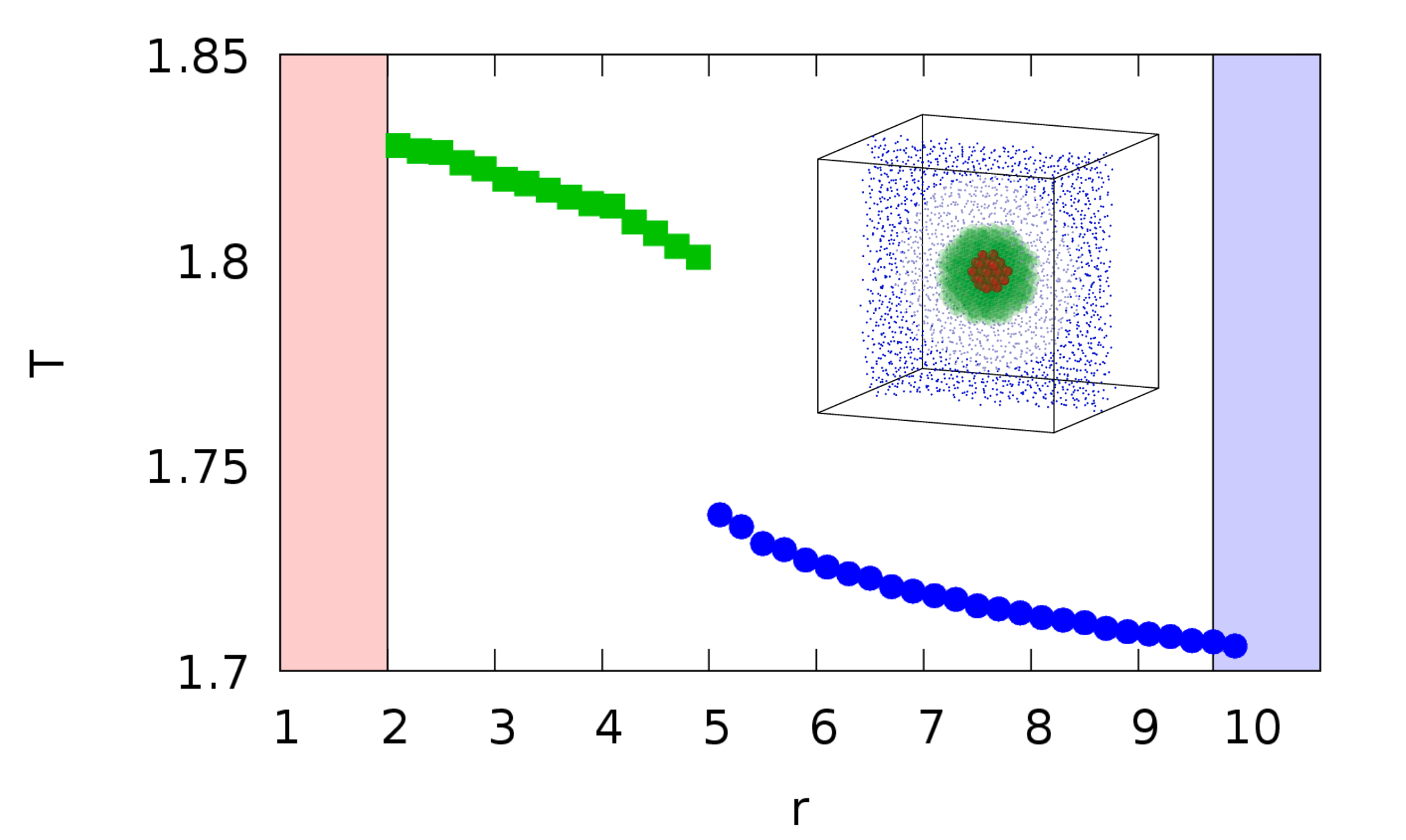} \\
      (a) & (b) & (c) 
    \end{tabular}
    \caption{(a) Snapshot of the simulation box in the \textit{external gradient} set-up, showing the thermostatting regions (red-hot, blue-cold). The solvent particles are represented as blue dots for visualization purposes, and the colloid as green spheres. (b) Temperature profile for a system with $\epsilon_{c}/\epsilon_{s}=20$, showing the temperature of the fluid (blue circles) and the colloid (green squares). (c) Temperature profile for the \textit{radial heat flux} system with $\epsilon_{c}/\epsilon_{s}=20$, showing the temperature of the fluid (blue circles) and the spherical particle (green squares), and the corresponding temperature jump. The inset shows a snapshot of the simulation box, highlighting the thermostatted core (red) and the surrounding solvent (blue dots).}
    \label{fig:snapshot}
\end{figure*}

\subsection{Test of the theoretical model using computer simulations}

We test the theoretical predictions using computer simulations of a colloid immersed in a WCA solvent at
reduced density, $\rho=0.8$ (see SI for details on the simulation setup and model). We also use the WCA potential to model the  
fluid-colloid interactions.
The
interactions inside the colloid are described with the spherically truncated and shifted Lennard-Jones potential with interaction strengths varying between 
$\varepsilon_{c}/\varepsilon_s=20 \cdots 100$, where $\varepsilon_s$ and $\varepsilon_c$ are the interaction strengths between solvent particles and the particles in the colloid, respectively.

We used two simulation set-ups to test the theoretical predictions. $G_{K}$, $\kappa^{s}$ and $\kappa^{c}$ were obtained using a \textit{radial heat flux} simulation set-up (see Fig. S1-c). The deformation of the temperature field due to the presence of the spherical particle was studied using an \textit{external gradient} set-up (see Fig.S1-a,b). In
both set-ups, the thermal gradients are simulated explicitly by setting hot and cold boundaries (see Fig. S1). The
\textit{external gradient} set-up was also employed to compute the thermal conductivity of the solvent and the Soret coefficient by computing the thermophoretic force associated with the displacement of the colloids attached with a harmonic spring to the geometric centre of the reservoirs shown in Fig. S1-a,b (see e.g. ref.\cite{bresme2022} for details on this method). The thermal conductivity and ITC were obtained from the heat flux using Fourier's law, $J_q = - \kappa \nabla T$  and the Kapitza relation, $J_q = G_K \Delta T$, where $\nabla T$ is the thermal gradient, and $\Delta T$ the temperature ``jump" at the colloid-solvent interface (see Fig. S1-c). Further simulation details are provided in the SI (see section 4).

\begin{figure*}[!ht]
    \centering 
    \begin{tabular}{ccc}
    \includegraphics[width=0.33\linewidth]{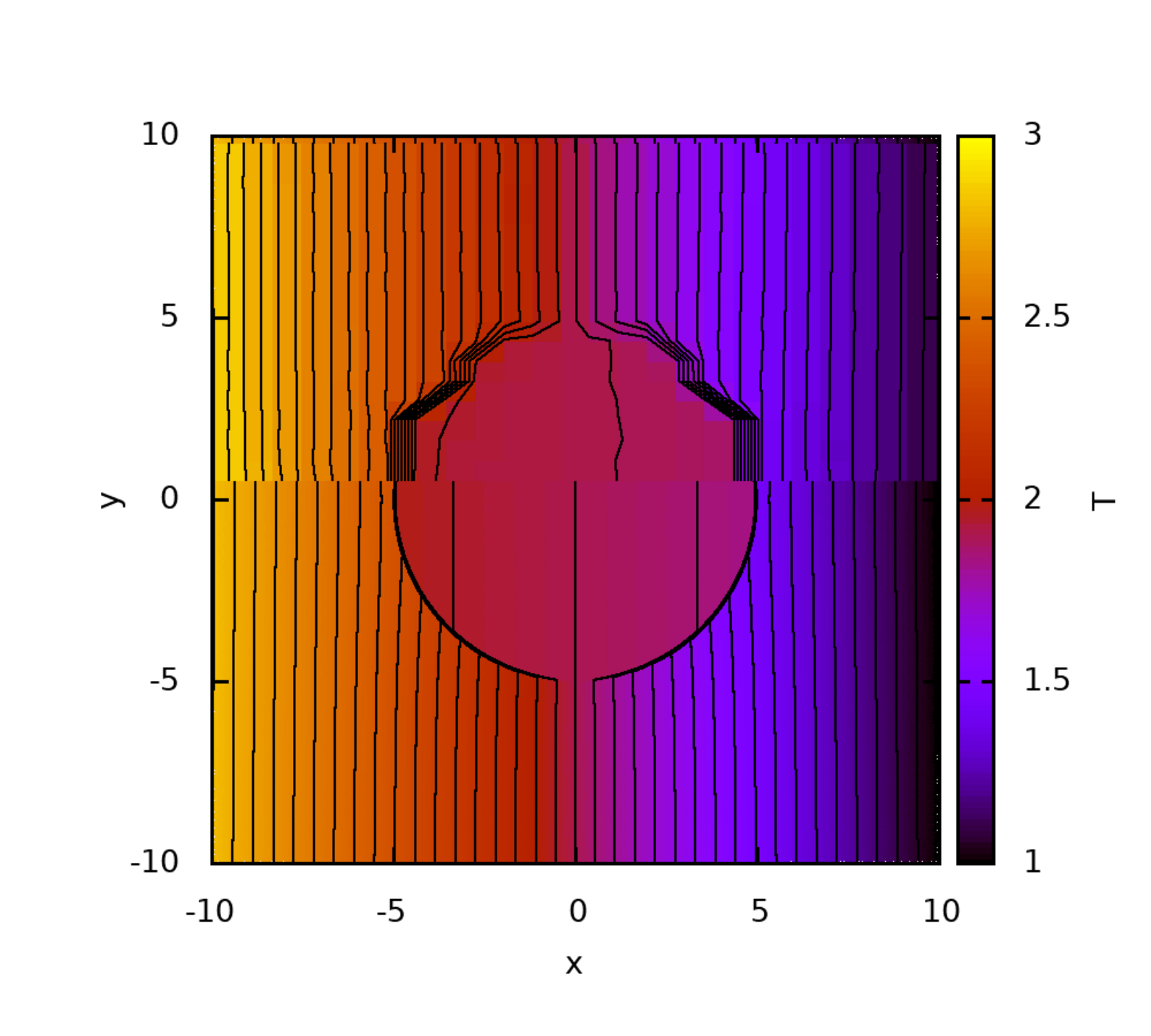} &
    \includegraphics[width=0.33\linewidth]{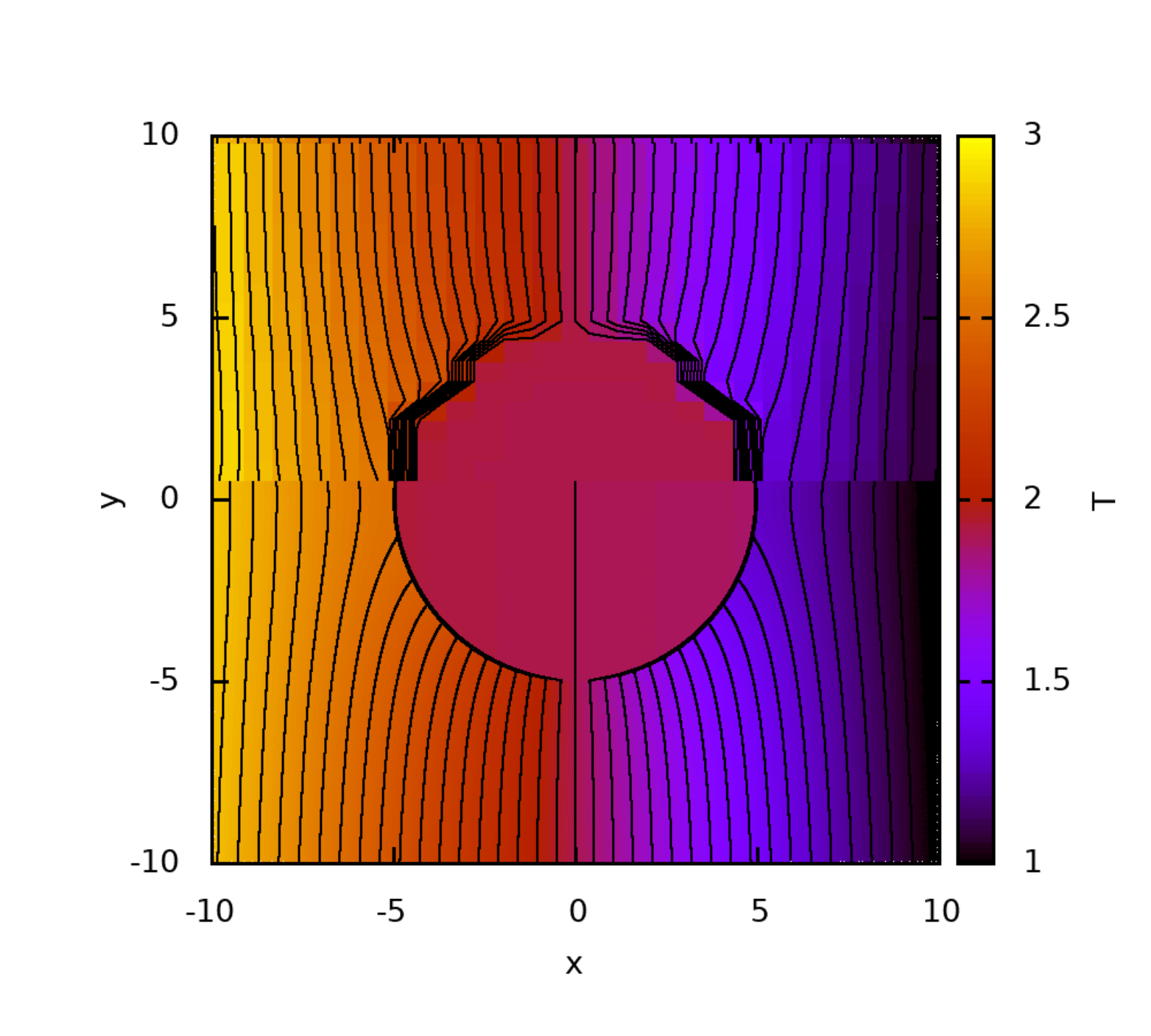} &
      \includegraphics[width=0.33\linewidth]{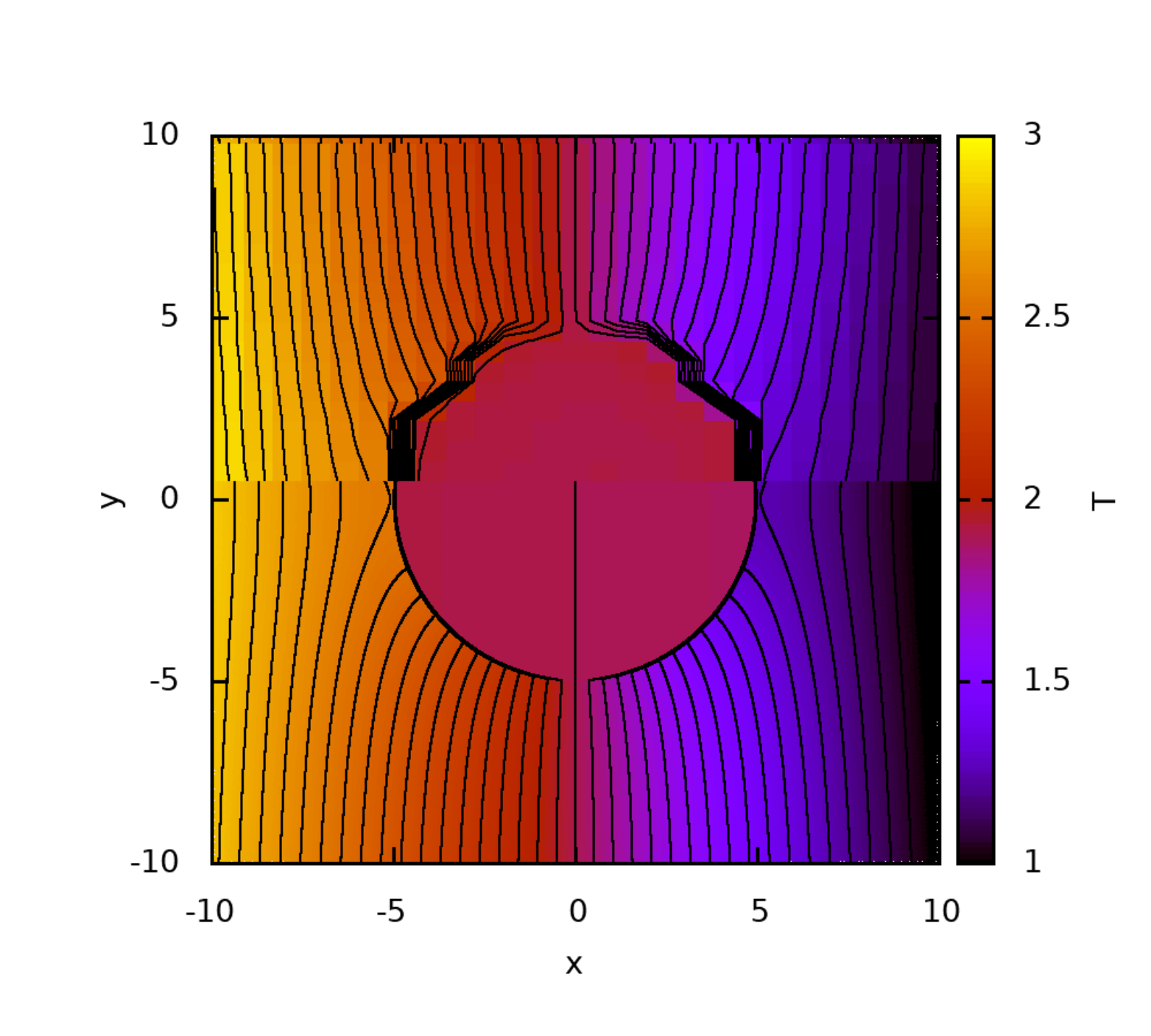} \\
   (a) & (b) & (c) 
    \end{tabular}
    \caption{Overlay of the NEMD simulation ($y>0$) and analytical results ($y<0$), for $\varepsilon_{c}/\varepsilon_{s}=$20, 60, 100 corresponding to three ITCs, $G_{K,LJ}$= 1.4 (a), 0.3 (b) and 0.11 (in Lennard Jones units, see SI). (c), and colloid thermal conductivities of  $\kappa^{c}=36$ (a), $\kappa^{c}=28$ (b) and $\kappa^{c}=20.5$ (c) (see SI for details regarding calculations of the thermal conductivities). All the results were obtained using $\kappa^{s}=6.87$, $T_0=1.9$ and $|\nabla T|=0.09$.}
    \label{fig:tmaps_overlay2}
\end{figure*}

First, we compare the predicted field around the spherical colloid against the simulated data using Eqs. (4) and (5) in the SI
, replacing $\alpha$ by $\alpha'$ to include ITC effects (see SI for details on the computation of the temperature field using NEMD simulations). We find good agreement between the predictions of the analytical solution and the NEMD results (see Fig. \ref{fig:tmaps_overlay2}), with some
deviations at the colloid surface, 
connected to the granularity of the colloid surface. Away from the interfacial region, the agreement of the temperature field is good, supporting the accuracy of Eqs. (4) and (5)
in the SI.

The impact of the ITC on the 
temperature field is also evident (c.f. Figs.\ref{fig:tmaps_overlay2}-a, b and c). The field features stronger deformation with decreasing ITC, evident at distances $>5$ solvent molecular diameters. Analytical and simulation results indicate that ITC effects are needed to describe temperature fields around colloids. 

\begin{figure}
    \centering
%    \begin{tabular}{cc}
 %       \includegraphics[width=0.5\linewidth]{dos-eps-converted-to.pdf}   &  
                    \includegraphics[width=1\linewidth]{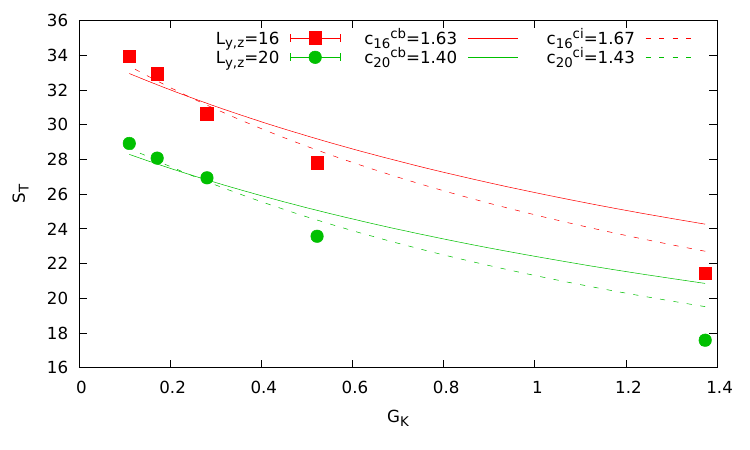}
     %         (a) & (b) \\
%    \end{tabular}
    \caption{Soret coefficient as a function of $G_K$ and system size. The solid lines represent fittings using the bulk fluid thermal conductivity (fitting parameter denoted by \textit{cb}. The dashed lines represent fittings to Eq. (\ref{eq:soret}) using the fluid interfacial thermal conductivity (fitting parameter denoted by \textit{ci}. In both cases, the colloid core thermal conductivity was used.}
    \label{fig:dos}
\end{figure}

We now discuss the theoretical predictions and simulated $S_T$.  
The $S_T$ computation
using the setup in Fig.S1-a 
involves restraining the translational motion of the colloid with a spring, which results in a hydrodynamic flow\cite{yang-2013,bresme2022} around the colloid and a system size-dependent $S_T$.
For a given system size, the $S_T$ 
is of the same order as the one obtained from the thermophoretic velocity of a freely drifting particle \cite{Lusebrink_2012}. 
We computed $S_T$ for system sizes L$_{y,z}= 16$ and 20. To test the numerical results against Eq. (\ref{eq:soret}), we used the 
colloid and solvent thermal conductivities corresponding to the bulk regions (see section 6 in the SI), the ITC obtained with the radial heat flux method (see Fig.S1-c), and the colloid radius, $R=5$. We set a constant fitting parameter in Eq. (\ref{eq:soret}), $c =4 \gamma_T \chi_{L_{y,z}}$, where $\chi_{L_{y,z}}$ takes into account the finite size of the simulation box and its impact on $S_T$.

Figure \ref{fig:dos} shows the dependence of the simulated Soret coefficients with $G_K$. The general dependence agrees with the theoretical results predicted by Eq. (\ref{eq:soret}) and those shown in Figure 1c, namely, $S_T$ decreases with increasing $G_K$.  We find that Eq. (\ref{eq:soret}) reproduces the trends of the simulated 
$S_T$, but some deviations are evident at large $G_K$. We note that we used the bulk solvent thermal conductivities in the theoretical calculation. However,  the density of the solvent next to the colloid differs significantly from the bulk density (see Fig. S2). Hence, we recalculated the solvent thermal conductivity, including only the region within one molecular diameter from the colloid surface. The thermal conductivity of this layer is lower than the bulk one (see Fig. S6 in the SI), and it leads to a  better agreement between the simulation data and the theory (see Fig. \ref{fig:dos}).

\section{Final remarks}

We have extended previous theories of the Soret effect\cite{Wurger2008,Giddings1995,Gaspard2019,Bickel2014}, and presented the analytical solution to the temperature field around a spherical colloid, including the Kapitza resistance, to account for the temperature discontinuity at the colloid-solvent interface. In our formulation, the  Soret coefficient varies with the deformation of the thermal field around the colloid, the thermal conductivities of colloid and solvent, and the ITC. The resulting equation includes corrected thermal conductivity terms to account for the ITC, $G_K$. We show that the  Soret coefficient depends strongly on ITC, for high colloid thermal conductivities ($\kappa_c \gg \kappa_s$), when $G_K <100-300$ W/(K m$^2$). These conditions can be found in gold nanoparticle suspensions. For $\kappa^c \sim \kappa^s$ (e.g. polymer nanoparticles), the impact of $G_K$ is small, and for $G_K \rightarrow \infty$, we recover previous theories, which ignored $G_K$ effects. We have verified the analytical equations using non-equilibrium molecular dynamics simulations of coarse-grained models of colloids immersed in a solvent. The simulated temperature fields agree with the ``continuum" theory, and the Soret coefficient varies with $G_K$ as predicted theoretically. 

Our work highlights the importance of non-equilibrium effects associated to explicit temperature gradients. Recently~\cite{Braun-prl-2023} it was suggested that thermophoresis is dominated by fluctuations at Peclet number (small thermal gradients), $Pe = R S_T \nabla T <1$, and by non-equilibrium transport for $Pe>1$. The system investigated here is consistent with the $Pe >1$ regime, where interfacial thermal gradients determine the thermophoretic motion. Furthemore, we find this non-equilibrium transport to dominate in a situation where the colloid is not much larger than the boundary layer.

Our work highlights the importance of the ITC in the thermophoresis of colloidal particles. This contribution has been ignored before, but as demonstrated here, it influences the temperature field around colloids and the thermophoretic force. 
We anticipate that the ideas presented here will be helpful in advancing the description of thermophoresis, a very complex non-equilibrium coupling effect with potential applications to colloidal trapping, nanofluidics, and the design of analytical devices.

\begin{acknowledgements}

We thank the Leverhulme Trust for grant RPG-2018-384 and the Imperial College High-Performance Computing Service for providing computational resources.

\end{acknowledgements}

\subsection{References}
\bibliography{refs}

\end{document}

\preprint{APS/123-QED}

\title{Supplementary Information: The impact of the interfacial thermal resistance on colloidal thermophoresis}% Force line breaks with \\

\author{Juan D. Olarte-Plata}
\email{j.olarte@imperial.ac.uk}
%  \altaffiliation[Also at ]{Physics Department, XYZ University.}%Lines break automatically or can be forced with \\
\author{Fernando Bresme}%
 \email{f.bresme@imperial.ac.uk}
\affiliation{Department of Chemistry, Imperial College London}
\date{\today}% It is always \today, today,

\begin{abstract}

\end{abstract}
\maketitle

\section{1.~Relevant equations for the temperature profiles around
colloids in the absence of interfacial thermal conductance effects}
For spherical colloids in an external temperature field $\nabla T$, of magnitude $|\nabla T|$ in the $\vec{x}$ direction, the temperature profile (in polar coordinates), far away from the particle is:

\begin{equation}
    T^{ext} = T_0 + |\nabla T| \vec{x} = T_0  +  |\nabla T|  r \cos \theta
    \label{eq:1}
\end{equation}

\noindent
where $T_0$ is a reference temperature far from the colloid. See the main text for definitions of the symbols below. Near the colloid, the solvent, $T^{s}(R, \theta)$ and colloid, $T^{c}(R, \theta)$ profiles fulfill the boundary conditions:

\begin{eqnarray}
    T^{s}(R, \theta) & = & T^{c}(R, \theta) \label{eq:boundary1} \\
    \left. \kappa^{s} \dfrac{\partial T^{s}(r, \theta)}{\partial r} \right|_{r=R} & = & \left. \kappa^{c} \dfrac{\partial T^{c}(r, \theta)}{\partial r} \right|_{r=R} \label{eq:boundary2} 
\end{eqnarray}

Following references \cite{Giddings1995,Bickel2014,Gaspard2019}, the solution of the Laplace equation is:
\begin{eqnarray}
T^{s}(r, \theta) & = & T_{0} + |\nabla T| r \cos \theta \left[1+\alpha\left(\dfrac{R}{r}\right)^3\right] \label{eq:4} \\
T^{c}(r, \theta) & = & T_{0} + |\nabla T| r \cos \theta \left[1+\alpha \right] \label{eq:5}
\end{eqnarray}
\noindent
where $R$ is the colloid radius. 
%The parameter  $\alpha$ quantifies the thermal conductivity contrast between the colloid and the fluid. 
An explicit equation for $\alpha$ follows from Eq. (\ref{eq:boundary2}) above,

\begin{equation}
    \alpha = \dfrac{\kappa^{s}-\kappa^{c}}{2\kappa^{s}+\kappa^{c}}
    \label{eq:6}
\end{equation}

\section{2.~Derivation of the equations discussed in the main text} 

We follow the approach of Würger to describe the relation between the surface stress and the Marangoni force for a freely drifting particle. The temperature gradient on the surface of the particle can be written as \cite{Wurger2008}:

\begin{equation}
    \left. \nabla_{\parallel}T = \left( \dfrac{1}{r} \dfrac{\partial T(r, \theta)}{\partial \theta} \right) \right |_{r=R} =   -|\nabla T| \sin \theta (1 + \alpha) \mathbf{t}
\end{equation}

\noindent where $\mathbf{t}$ is the tangential vector at the surface. The Marangoni force is given by \cite{Wurger2008}:

\begin{equation}
    \nabla_{\parallel} \gamma = \gamma_{T} \nabla_{\parallel}T
\end{equation}
\noindent
\noindent where the Marangoni parameter is defined as the derivative of the interfacial free energy, $\gamma$, with respect to the temperature, $\gamma_{T}=d \gamma / dT$. The dissipative part of the stress tensor, $\mathbf{\sigma'}$, is given by \cite{Wurger2008}:

\begin{eqnarray}
    \sigma'_{rr} & = & 2 \eta \dfrac{\partial\hat{v}_{r}}{\partial r} \\
    \sigma'_{r\theta} & = & \eta \left(\dfrac{1}{r} \dfrac{\partial \hat{v}_r}{\partial \theta} + \dfrac{\partial \hat{v}_{\theta}}{\partial r} - \dfrac{\hat{v}_{\theta}}{r}\right) \label{eq:tangentialstress}
\end{eqnarray}

\noindent where $\eta$ is the viscosity of the fluid. $\hat{v}_{r}$ and $\hat{v}_{\theta}$ are the radial and tangential components of the fluid velocity field near the particle, in the particle frame of reference.

For a particle drifting with velocity $\mathbf{u} = u \hat{x}$ (due to the thermophoretic force), the solution to the Stokes equation for an incompressible fluid, $\eta \nabla^2 \hat{\mathbf{v}} = \nabla P$, can be written as \cite{Wurger2008}:

\begin{eqnarray}
    \hat{v}_{r} & = & -u \cos \theta \left(1-\dfrac{R^3}{r^3} \right) \\
    \hat{v}_{\theta} & = & u \sin \theta \left(1+\dfrac{R^3}{2r^3} \right) 
\end{eqnarray}

For this case, the tangential component of the stress tensor evaluated at the particle surface is:

\begin{equation}
    \left. \sigma'_{r\theta} \right|_{r=R} = \dfrac{-3 \eta u \sin \theta}{R}
\end{equation}

The thermophoretic velocity of the particle can be found by balancing the tangential component of the stress tensor with the Marangoni force, $\sigma'_{\theta} \mathbf{t} = -\gamma_{T} \nabla_{\parallel} T$ \cite{Wurger2008}, giving:

\begin{equation}
    \dfrac{-3 \eta u \sin \theta}{R} = \gamma_{T} |\nabla T| \sin \theta (1 + \alpha)
\end{equation}

The drift velocity of the particle can thus be written as:

\begin{equation}
\label{eqn:therm-veloc}
    u = -\dfrac{\gamma_{T} R (1+\alpha')}{3 \eta} |\nabla T| 
\end{equation}
\noindent

\section{3.~Forces on a fixed particle}
In our simulations, the translational motion of the nanoparticle is restrained in the middle plane between the hot and cold thermostats.
This physical situation corresponds to that of a fixed particle and results in a velocity field distinctively different from the one obtained with the freely drifting particle. However, the corresponding friction force is the same except for a constant. Below, we derive the relevant equations for this case.

The total force on the particle is given by~\cite{Landau1987Fluid}
\begin{equation}
    F_{tot} = \oiint\limits_{S} (-p \cos \theta + \sigma_{rr}' \cos \theta -  \sigma_{r \theta}'  \sin \theta) dA
\end{equation}

\noindent where $p$ is the pressure, and $\sigma_{ij}'$ are components of the stress tensor.

The general solution for the fluid velocity around a spherical particle is given by \cite{Landau1987Fluid}:

\begin{eqnarray}
    v_{r} & = & u_{0} \cos \theta \left(1-\dfrac{2a}{r}+\dfrac{2b}{r^3} \right) \\
    v_{\theta} & = & -u_{0} \sin \theta \left(1-\dfrac{a}{r}-\dfrac{b}{r^3} \right) 
\end{eqnarray}

\noindent where the constants $a$ and $b$ are determined from the boundary conditions. For instance, a freely drifting particle with \textit{stick} boundary conditions implies that at $r=R$, ${v}_{r}=0$ and ${v}_{\theta}=0$. The solution is given by $a=3R/4$ and $b=R^3/4$, which implies:

\begin{eqnarray}
    \hat{v}_{r} & = & u_{0} \cos \theta \left(1-\dfrac{3R}{2r}+\dfrac{R^3}{2r^3} \right) \\
    \hat{v}_{\theta} & = & -u_{0} \sin \theta \left(1-\dfrac{3R}{4r}-\dfrac{R^3}{4r^3} \right)     
\end{eqnarray}

For the fixed particle, the fluid velocity can be written as \cite{Morthomas2010}:

\begin{eqnarray}
    v_{r} & = & u_{0} \cos \theta \left(\dfrac{R}{r}-\dfrac{R^3}{r^3} \right) \\
    v_{\theta} & = & - \dfrac{u_{0} \sin \theta}{2}  \left(\dfrac{R}{r}+\dfrac{R^3}{r^3} \right) 
\end{eqnarray}

In this case, $u_{0}$ represents the velocity of the fluid at the particle's surface. With these expressions for the velocity field, we arrive to:

\begin{eqnarray}
    \sigma_{rr}'|_{_{r=R}} & = & 4 \eta \dfrac{u_0}{R} \cos \theta \\
    \sigma_{r\theta}'|_{_{r=R}} & = & 3 \eta \dfrac{u_0}{R} \sin \theta
\end{eqnarray}

We note that the tangential stress, $\sigma_{r\theta}'$, has the same solution as the freely drifting particle with surface forces (see section 2 in the SI). This justifies using the approach proposed in Ref. \cite{Wurger2008} when equating the tangential stress to the Marangoni force.

The corresponding contributions to the total force are given by:

\begin{eqnarray}
    F_{\sigma_{rr}'} & = & \dfrac{16 \pi}{3} \eta R u_0 \\
    F_{\sigma_{r\theta}'} & = & -8 \pi \eta R u_0
\end{eqnarray}

The pressure contribution to the total force can be found from the following expression \cite{Wurger2008}:

\begin{equation}
    P = P_0 + 2 \alpha \cos \theta \dfrac{\eta u_0 R}{r^2}
\end{equation}

\noindent where $\alpha=1/2$ corresponds to the solution for the fluid velocity field around the fixed particle. The pressure contribution to the total force thus reads:

\begin{equation}
    F_{P} = -\dfrac{4 \pi}{3} \eta R u_0
\end{equation}

The total force on the particle is thus given by:

\begin{equation}
    F_{total} = F_{\sigma_{rr}'} + F_{\sigma_{r\theta}'} + F_{P} = -4 \pi \eta R u_0
\end{equation}

From the previous equation, we note that the friction coefficient of the fixed particle with surface forces has a factor of $4 \pi$, as discussed in the main text.

\section{4.~Simulation details}

The fluid, with reduced density $\rho= \rho_n \sigma^3= 0.8$ ($\rho_n$ is the number density in particles/m$^{-3}$) was described using the WCA model, {\it i.e.} a Lennard-Jones 12-6 potential, $4\varepsilon_s \left[(\sigma_s/r)^{12}-(\sigma_s/r)^{6}\right]$, with a cut-off radius corresponding to the minimum of the potential, \textit{i.e.} $r_c=r_0=2^{1/6} \sigma$, with $\epsilon_{s}=1.0$. $\sigma$ is the diameter of the solvent and the particles inside the colloid. We used reduced units, $r= r_{SI}/\sigma$, $T = k_B T_K / \varepsilon_s$, where $r_{SI}$ is the distance in SI units, $\varepsilon_s$ is the interaction strength between solvent particles and $T_K$ the temperature in Kelvin. in these units the ITC is defined as, $G_{K,LJ}= G_K \sigma^2/k_B \sqrt{m \sigma^2/\varepsilon_s}$. All the solvent-particle interactions were computed using the WCA model. The interactions between the particles inside the colloids were described using a strongly attractive Lennard-Jones potential with interaction strength~$\varepsilon_s$ and a cut-off radius $r_c=2.5 \sigma$.

\begin{figure}
    \centering
    \begin{tabular}{ccc}
        \hspace{0.75cm} \includegraphics[width=0.8\linewidth,valign=c]{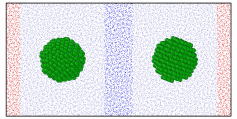} (a) \\
       \includegraphics[width=0.9\linewidth,valign=c]{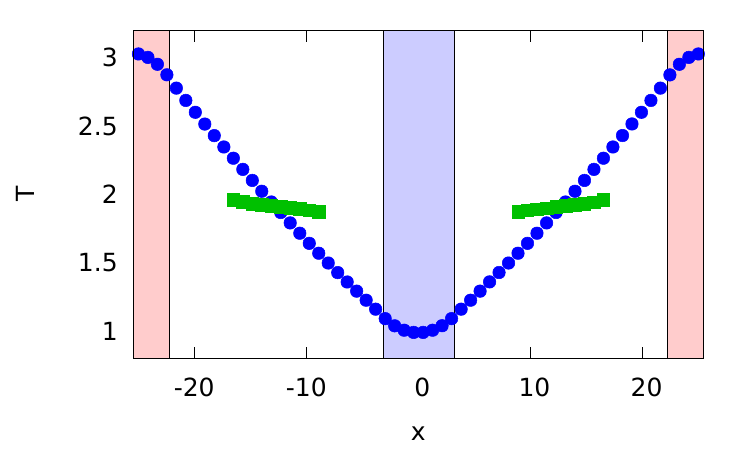} (b) \\
      \includegraphics[width=0.9\linewidth,valign=c]{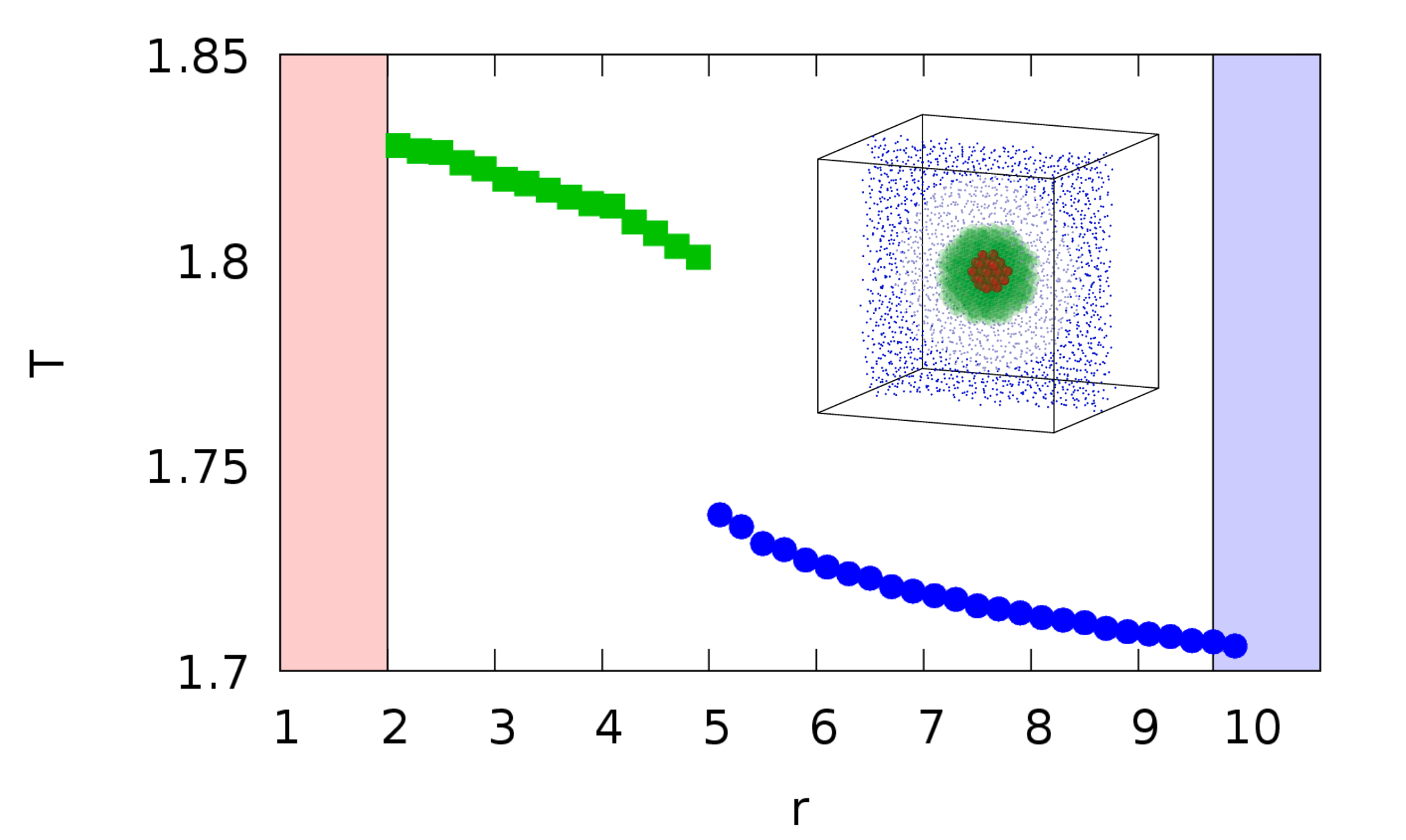} (c) \\
      %(a) & (b) & (c) 
    \end{tabular}
    \caption{(a) Snapshot of the simulation box in the \textit{external gradient} set-up, showing the thermostatting regions (red-hot, blue-cold). The solvent particles are represented as blue dots for visualization purposes, and the colloid as green spheres. (b) Temperature profile for a system with $\epsilon_{c}/\epsilon_{s}=20$, showing the temperature of the fluid (blue circles) and the colloid (green squares). (c) Temperature profile for the \textit{radial heat flux} system with $\epsilon_{c}/\epsilon_{s}=20$, showing the temperature of the fluid (blue circles) and the spherical particle (green squares), and the corresponding temperature jump. The inset shows a snapshot of the simulation box, highlighting the thermostatted core (red) and the surrounding solvent (blue dots).}
    \label{fig:snapshot}
\end{figure}

The simulations were run with a timestep $\delta t= 0.0025$ in Lennard Jones units, and using LAMMPS \cite{lammps}. After $10^4$ timesteps in the NVT ensemble with $T=1.0$, the thermostats (see Figure~\ref{fig:snapshot}) were activated. The first $10^5$ timesteps were discarded before sampling for an additional $10^7$ timesteps. The results reported in our work were obtained using 20 independent replicas, starting from random atomic velocities and colloid orientations. 
To perform the simulations using the two setups (see Fig. \ref{fig:snapshot}), we generated cubic (\textit{radial heat flux}) or rectangular (\textit{external gradient}) simulation boxes with volume  $V = (16 a_0) \times (16 a_0) \times (16 a_0)$ or $V = (32 a_0) \times (16 a_0) \times (16 a_0)$, where $a_0 = 2^{2/3} \sigma$ is the FCC lattice parameter of a solid with $\rho^*=1.0$. To investigate finite-size effects in the computations of the Soret coefficient, we also simulated systems with $L_{y,z}=20 a_0$ for the \textit{external gradient} set-up. In all cases, $L_x = 2L_{y,z}$. To set a temperature gradient, hot and cold thermostatting regions were defined at the center and edges of the simulation box. The thermostatting regions correspond to a spherical core, $r<2.0\sigma$, and shell, $R>9.7\sigma$ (\textit{radial heat flux}), or rectangular slabs of width $4a_0$ (\textit{external gradient}). Within these regions, the velocities of the fluid particles were rescaled to the target temperatures every 100 timesteps. The center-of-mass momentum was also removed with the same frequency to prevent a drift in the translational motion of the atoms. The \textit{external gradient} configuration generates two temperature gradients in opposite directions. A spherical particle, cut from an FCC lattice with density $\rho=1.0$ and radius $R=5\sigma$, was placed in the center of each compartment and attached with harmonic potential, given by $U(r) = 0.5 \ k (r-r_0)^2$, with force constant $k=10^3$. Snapshots of the simulation box and typical temperature profiles are shown in Fig. \ref{fig:snapshot}. All the visualizations were done using OVITO \cite{ovito}.

The temperature fields were calculated by dividing the simulation box in cubic voxels of volume  $\sim(0.79 \sigma)^3$, where the local temperature was sampled. To exploit the symmetry of the problem and improve sampling, the data was re-binned (using a weighted average) to $(x,r)$ coordinates, where $r=\sqrt{y^2+z^2}$. The bin size for the $r$ coordinate was $\sim0.56\sigma$. The contours were generated using a linear interpolation, with increasing levels of 0.05.

\section{5.~Density profiles of the fluid}

\begin{figure}[!ht]
    \centering
    \includegraphics[width=\linewidth]{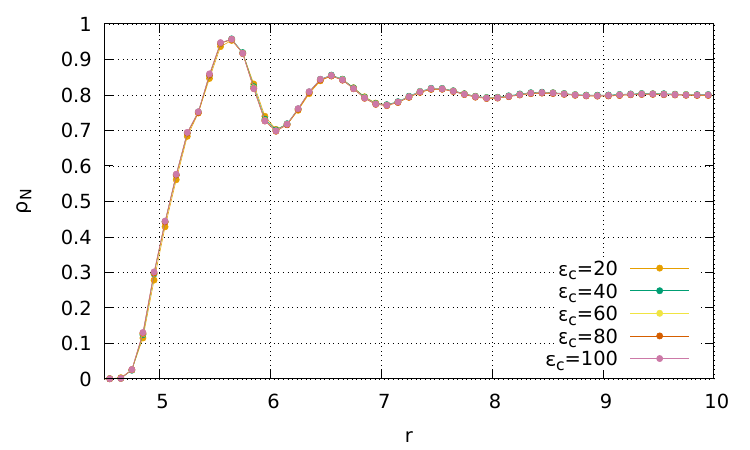}
    \caption{Solvent radial density profiles around the colloid, as a function of the colloid interaction strength. The profiles were obtained at equilibrium conditions and $T=2.0$.}
    \label{fig:density_profiles}
\end{figure}

Fig. \ref{fig:density_profiles} shows the radial density profile of the fluid as a function of radial distance to the center of geometry of the colloid and the colloid interaction strength, obtained under equilibrium conditions at $T=2.0$. The results in Fig. \ref{fig:density_profiles} show that the colloid interactions do not impact the density profiles of the solvent surrounding the colloid.

\section{6.~Calculation of the colloid-solvent Interfacial Thermal Conductance, $G_K$}

To compute the interfacial thermal conductance $G_K$ we performed simulations targeting the highest and lowest interfacial temperatures obtained in the simulations of the Soret coefficient (see Fig. \ref{fig:snapshot}). Our simulations also consider the rectification of the $G_K$, namely, the solvent is hotter or colder on both sides of the colloid (see Fig. \ref{fig:snapshot}).

\begin{figure}
    \centering
        \includegraphics[width=1\linewidth]{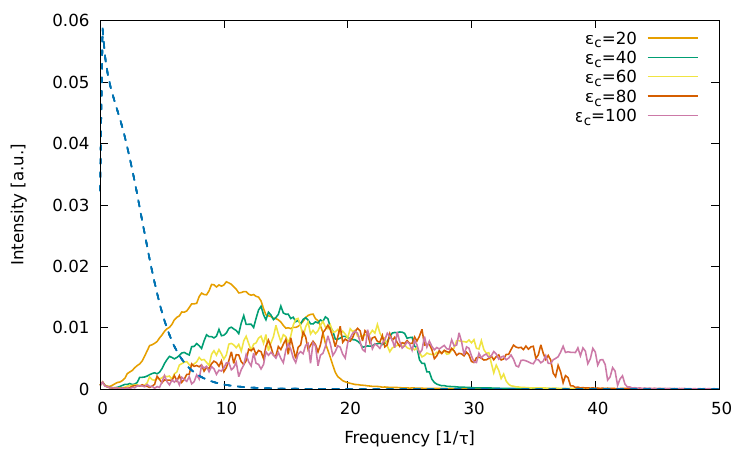}   
    \caption{(a) VDoS of the colloid and solvent as a function of $\varepsilon_c$. The fluid VDoS is represented with a blue dashed line.}
    \label{fig:dos2}
\end{figure}

\begin{figure}[!ht]
    \centering
    \includegraphics[width=\linewidth]{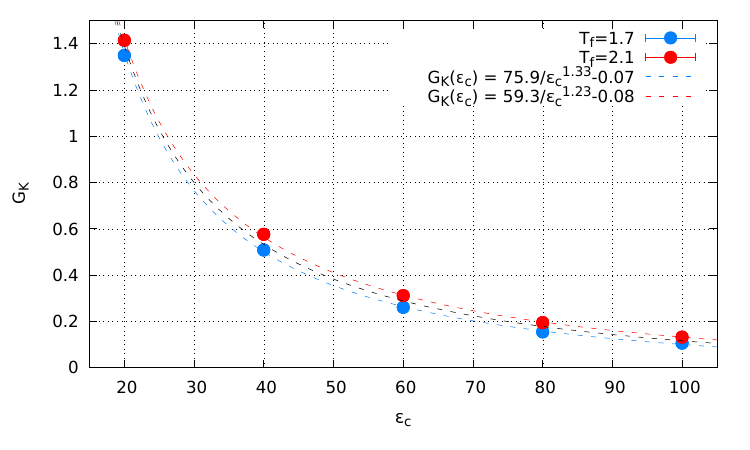}
    \caption{Interfacial thermal conductance as a function of the colloid internal interaction strength for different bulk fluid temperatures. The dashed lines indicate fittings to a function of the form $G(\epsilon_{c}) = a/\epsilon_{c}^{b}+d$. }
    \label{fig:conductance}
\end{figure}

We now discuss the dependence of the Soret coefficient on the ITC. To address this point, we vary $G_K$, keeping the solvent-particle interaction strength constant and approximately the same interfacial temperature; hence, $\gamma_T$ does not change with $G_K$. This notion is supported by the lack of dependence of the colloid solvation structure with $\varepsilon_c$ (see Fig. \ref{fig:density_profiles}). By modifying the colloid particle-particle interactions, we isolate $G_K$ and surface tension effects. This leads to a systematic change of the colloid vibrational density of states (VDoS) relative to the VDoS of the solvent. By changing the overlap of the two VDoS we achieve higher or lower $G_K$, and we can systematically test Eq. (9) in the main paper. %(\ref{eq:soret}). 
Figure \ref{fig:dos2}
%\ref{fig:dos}-a 
shows the impact of the $\varepsilon_{c}$ on the overlapping of the colloid and solvent Vibrational Density of States (VDoS). The ITC decreases with increasing $\varepsilon_c$ due to a shift of the colloid VDoS to higher frequencies as $\varepsilon_{c}$ increases (see Fig. \ref{fig:conductance}). 

The VDoS was computed as the Fourier transform of the velocity autocorrelation function for the fluid and solid particles:

\begin{equation}
 VDoS_{j,k}(\omega) =  \int\limits_{0}^{\infty} \frac{\left\langle {v}_{j,k}(0)  {v}_{j,k}(t) \right\rangle }{\left\langle {v}_{j,k}(0)  {v}_{j,k}(0) \right\rangle }\exp(- 2 \pi i \omega t) \mathop{dt}
 \label{eq:dos}
\end{equation}
\noindent
where $\left\langle {v}_{j,k}(0) \cdot {v}_{j,k}(t) \right\rangle$ is the velocity autocorrelation function of particle type $j$ and component $k$. We used a simulation box equivalent to Fig. S1a with no temperature gradient and average temperature $T=2.0$.  The velocity autocorrelation function for fluid and colloid particles was calculated for 100 windows of 2500 timesteps each (using $\delta t=0.001$) and averaged over ten replicas. All the colloid particles were included in the calculation.

\begin{figure*}[!ht]
    \centering
    \begin{tabular}{ccc}
        \rotatebox{90}{\hspace{2.cm} $\epsilon_{c} = 20$} & \includegraphics[width=0.45\linewidth]{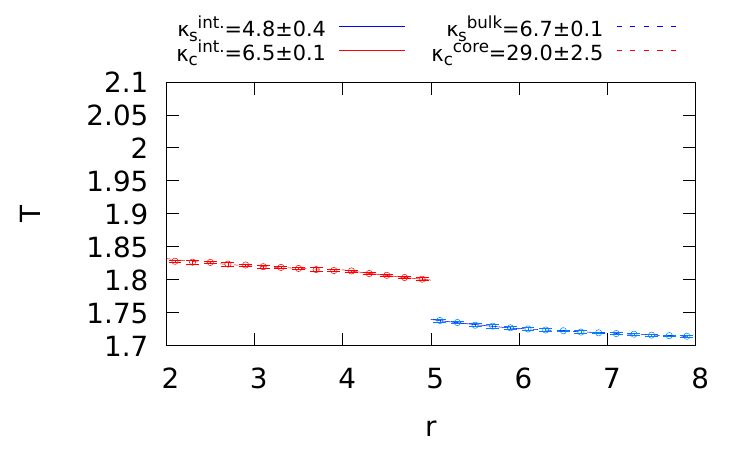}   &  
        \includegraphics[width=0.45\linewidth]{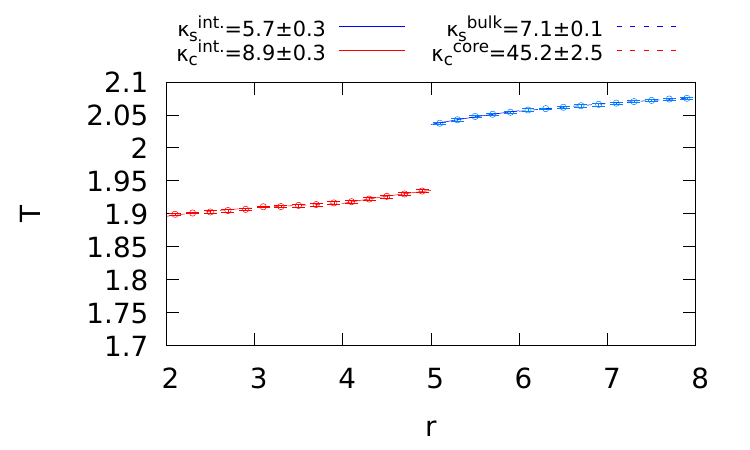}    \\
        \rotatebox{90}{\hspace{2.cm} $\epsilon_{c} = 40$} & \includegraphics[width=0.45\linewidth]{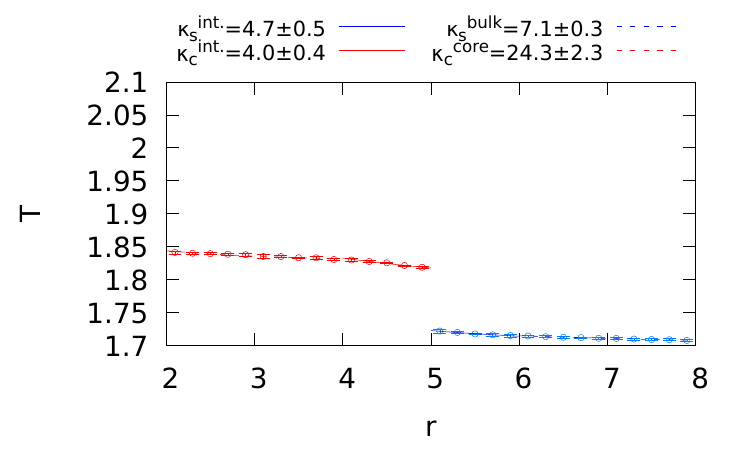}   &  
        \includegraphics[width=0.45\linewidth]{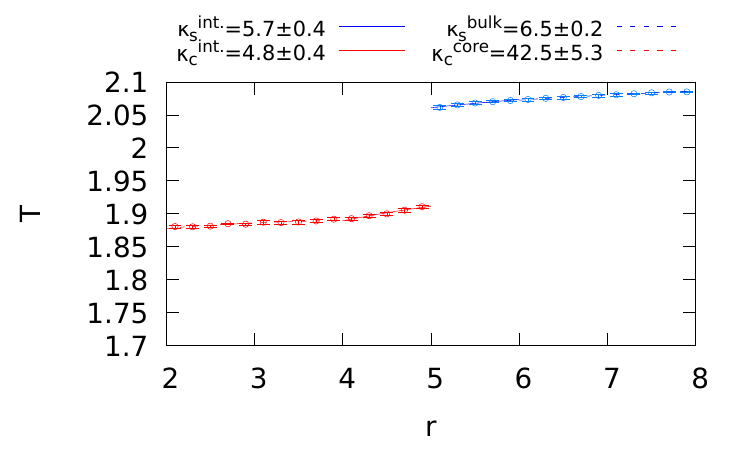}    \\
        \rotatebox{90}{\hspace{2.cm} $\epsilon_{c} = 60$} & \includegraphics[width=0.45\linewidth]{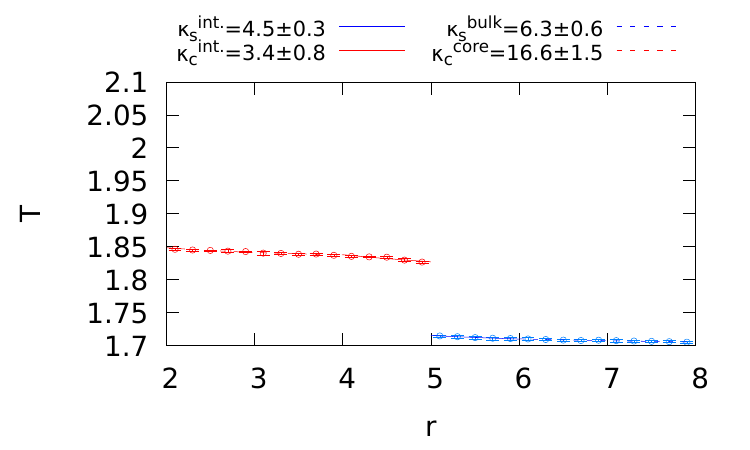}   &  
        \includegraphics[width=0.45\linewidth]{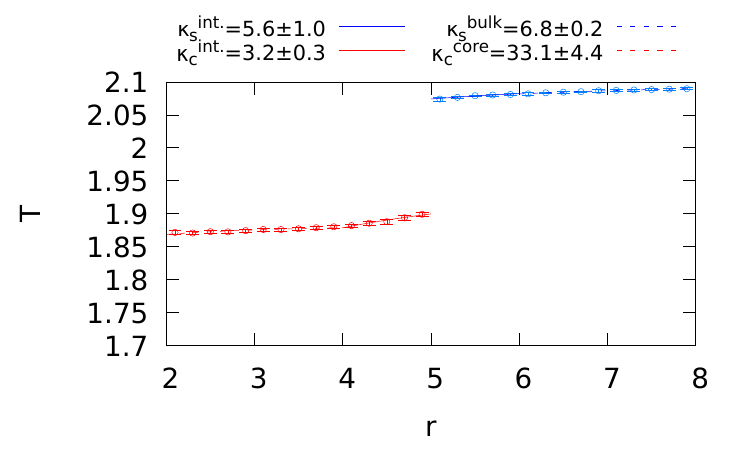}    \\
    \end{tabular}
    \caption{Temperature profiles in the \textit{radial heat flux} configuration, for different boundary conditions and colloid interaction strengths. The lines represent the fitting used to calculate the thermal gradient and thermal conductivities. $\kappa_{\alpha}^{\beta}$ refers to the colloid ($\alpha=c$) or solvent ($\alpha=s$) thermal conductivity obtained using the $\beta=$``bulk" for the solvent, $\beta=$``core" for the colloid or interfacial region $\beta=$``int" for the solvent. Error bars represent the standard deviation calculated from the individual temperature profiles of 10 replicas.}
    \label{fig:radial_profiles}
\end{figure*}

We compute the interfacial thermal conductance of uniform nanoparticles as a function of their internal interaction strength using the \textit{radial heat flux} simulation set-up (see Fig. \ref{fig:snapshot}). The temperature of the colloid core is thermostatted to a target temperature $T_c=1.9$. The temperature of the bulk solvent is varied between $T_s=1.7$ and $T_s=2.1$ to account for the colloid-solvent contact on the hot and cold regions (see Fig. \ref{fig:snapshot}). For the calculation of the temperature ``jump", we fitted the temperature profiles using the heat diffusion equation near the interfacial region (see Fig. \ref{fig:radial_profiles}). The range of radial distance for the colloid was restricted to $r^*_{c}=[4.0,5.0]$, while for the solvent we used the interval $r_{s}=[5.0,6.0]$.

\clearpage

\section{6.~Thermal conductivity of the solvent and colloid}

\begin{figure}[h]
    \centering
    \includegraphics[width=\linewidth]{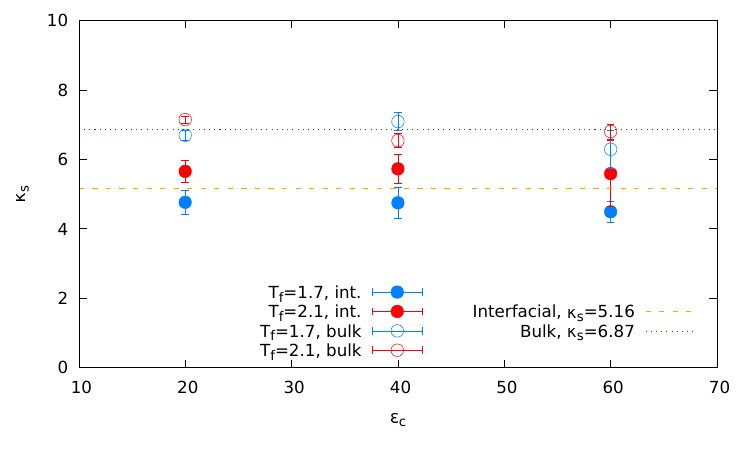}
    \caption{Thermal conductivity of the solvent as a function of the colloid interaction strength and the temperature. The upper values were obtained using the temperature gradient in the bulk region and the lower values were obtained using the interfacial temperature gradient (see Fig. \ref{fig:radial_profiles} for definitions of bulk and interfacial regions). Red and blue points represent the thermal conductivity at different temperatures, targeting the temperatures at the hot or cold side of the colloid surface (see Figure \ref{fig:snapshot}). The dashed horizontal lines represent the average thermal conductivities for each temperature and varying $\epsilon_c$.}
    \label{fig:kf}
\end{figure}

The thermal conductivity of the solvent was calculated using the {\it radial heat flux} simulation set-up (see Fig. \ref{fig:snapshot}). Due to the decreasing heat flux with increasing particle interaction strength, we focused on the range $\varepsilon_{c}=20-60$. Fig. \ref{fig:kf} shows the thermal conductivity as a function of the fitting region and the boundary condition of the fluid. The thermal conductivity does not feature a significant variation with
either temperature or particle interaction strength $\varepsilon_c$, 
Hence, to plot the theoretical curves in Fig. 3 in the main text, 
we used the thermal conductivities {$\kappa_s^{bulk}=6.87$} and {$\kappa_{s}^{int.}=5.16$} (see dashed horizontal lines in Fig. \ref{fig:kf}).

\begin{figure}[b]
    \centering
    \includegraphics[width=\linewidth]{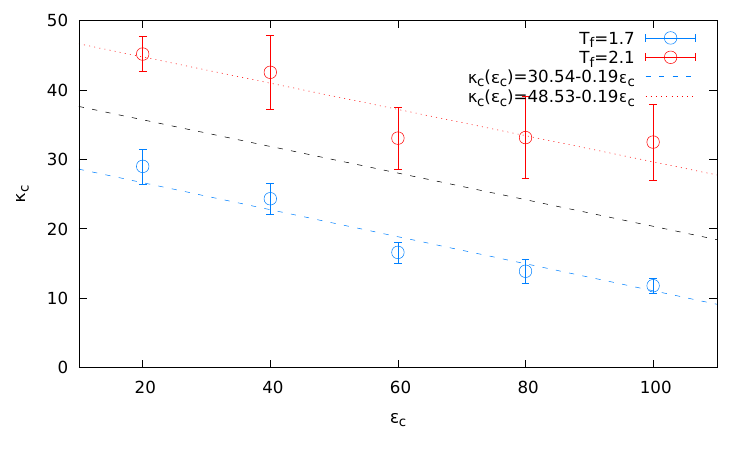}
    \caption{Thermal conductivity of the colloid as a function of the colloid interaction strength for different temperatures. The dashed lines indicate fittings to a linear function. The black dashed lines represent the average of both high and low-temperature lines.}
    \label{fig:kp}
\end{figure}

The thermal conductivity of the colloid (see Fig. \ref{fig:kp}) was calculated using the {\it radial heat flux} simulation set-up. We used Fourier's law by fitting the temperature gradient inside the colloid in an interval far from the colloid surface to avoid including interfacial effects (the interval $r_{s}=[2.0,4.0]$).

Error bars for the thermal conductivities represent the asymptotic standard error from the fitting of the temperature profiles (and their associated standard deviations) shown in Fig. \ref{fig:radial_profiles}.

% ccc
% The thermal conductivity of the colloid appearing in Eq. (\ref{eq:soret}) was computed by using the radial heat flux setup (see Fig. \ref{fig:snapshot}-c). We used Fourier's law by fitting the temperature gradient inside the colloid in an interval far from the colloid surface to avoid including interfacial effects (see section 6 in the SI for details). In the SI, we show in Fig. S5 that the colloid thermal conductivities decrease with increasing  $\varepsilon_c$. These thermal conductivities were used as input for Eq. (\ref{eq:soret}). The final element needed to test Eq. (\ref{eq:soret}) is the Soret coefficient. 
% ccc

\section{7.~ Estimation of the effective thermal conductivities of functionalized colloids}

We used the following approach to estimate the effective thermal conductivity of the functionalized colloids discussed in Fig. 1c in the main paper. The effective thermal conductivity of the colloid, $\kappa_{eff}$, can be estimated using the model introduced by Hasselman and Johnson~\cite{Hassleman1987}, which reduces to the expression derived by Maxwell for infinite interfacial thermal conductance. As an example we consider here a gold colloid with $\kappa_{Au} \sim$ 300 W/(K m), coated with an alkanethiol monolayer with  $\kappa_m \sim $ 0.45 W/(K m)~\cite{olarte-plata-acs-nano2022} at $\sim 300$~K. The interfacial thermal conductance Au-monolayer interface is $\sim$ 800 MW/ (K m$^2$)\cite{olarte-plata-acs-nano2022}. Following ref. \cite{Hassleman1987} the effective thermal conductivity of the colloid is given by:

\begin{equation}
    \kappa_{eff} = \kappa_m \dfrac{2 \left(\dfrac{\kappa_{Au}}{\kappa_m} - \dfrac{\kappa_{Au}}{R G_K} -1 \right) V_{Au} + \dfrac{\kappa_{Au}}{\kappa_m} + \dfrac{2 \kappa_{Au}}{R G_K} +2}{\left( 1- \dfrac{\kappa_{Au}}{\kappa_m} + \dfrac{\kappa_{Au}}{R G_K}  \right)V_{Au} + \dfrac{\kappa_{Au}}{\kappa_m} + \dfrac{2 \kappa_{Au}}{R G_K} +2 }
\end{equation}
\noindent
where $V_{Au}= (R/(R+d))^3$ is the volume fraction of the gold particle with radius $R$, relative to the total volume of the colloid, defined by the radius $R+d$, where $d$ is the thickness of the monolayer coating the gold core.  

Considering a colloid of radius $R=$250 nm and monolayers of thickness between 1 or 2 nm,  we obtain
the effective thermal conductivities 58.6 W/(K m) or 38.8 W/(K m), respectively. Ignoring the interfacial thermal conductance, $G_K \rightarrow \infty$, we find that the effective thermal conductivity is higher, 83.0 and 48.2 W/(K m) for monolayer thicknesses of 1 or 2 nm, respectively.

\section{8.~Soret coefficient vs. internal interaction strength and system size}

\begin{figure}[t]
    \centering
    \includegraphics[width=\linewidth]{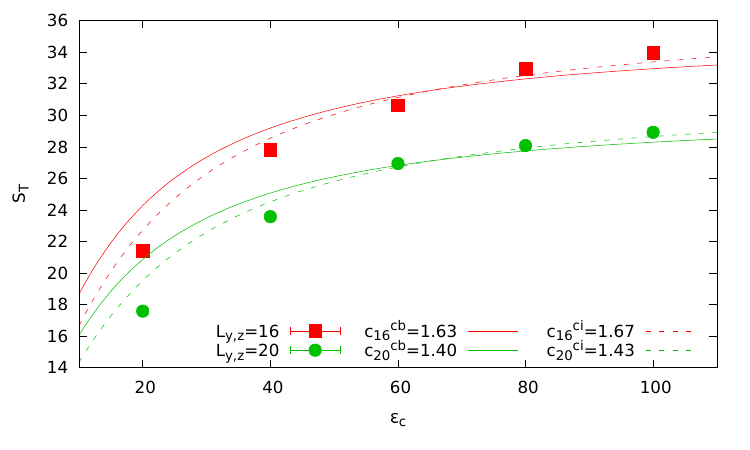} \\
    \caption{Soret coefficient as a function of the particle interaction, $\epsilon_{c}$, and system size. The symbols represent simulation results, and the lines fittings to the theoretical equations. See the discussion in the main text.}
    \label{fig:soret_size}
\end{figure}

Fig. \ref{fig:soret_size} shows the Soret coefficients obtained with the method discussed in section 2 above (see also discussion in the main paper pages 3 and 4). We considered different colloid interaction strengths as well as different system sizes. The Soret coefficient decreases with increasing system size (see Fig. \ref{fig:soret_size} and also the discussion on the truncation of the hydrodynamic field in reference \cite{bresme2022}).

\bibliographystyle{apsrev4-1}
\bibliography{refs}